\documentclass[conference, 9pt]{IEEEtran}
\usepackage{graphicx}
\usepackage{multirow}
\usepackage{epsfig}
\usepackage{url}
\usepackage{algorithmic}
\usepackage{algorithm}
\usepackage[numbers, sectionbib,comma, sort&compress,square]{natbib}

\author{\IEEEauthorblockN{Wei-Che Wang\IEEEauthorrefmark{1}, Zhuoqi Li\IEEEauthorrefmark{1}, Joseph Skudlarek\IEEEauthorrefmark{2}, Mario Larouche\IEEEauthorrefmark{2}, Michael Chen\IEEEauthorrefmark{2} and Puneet Gupta\IEEEauthorrefmark{1} }
\IEEEauthorblockA{\IEEEauthorrefmark{1}Department of Electrical Engineering, University of California, Los Angeles}
\IEEEauthorblockA{\IEEEauthorrefmark{2}Mentor Graphics}
}

\usepackage{amsmath}
\usepackage{gensymb}
\usepackage{makecell}

\include{conferenceList_full}
\title{UNBIAS PUF: A Physical Implementation Bias Agnostic Strong PUF}

\begin{document}
\maketitle
\begin{abstract}
The Physical Unclonable Function (PUF) is a promising hardware security primitive because of its inherent uniqueness and low cost. To extract the device-specific variation from delay-based strong PUFs, complex routing constraints are imposed to achieve symmetric path delays; and systematic variations can severely compromise the uniqueness of the PUF. In addition, the metastability of the arbiter circuit of an Arbiter PUF can also degrade the quality of the PUF due to the induced instability. In this paper we propose a novel strong UNBIAS PUF that can be implemented purely by Register Transfer Language (RTL), such as verilog, without imposing any physical design constraints or delay characterization effort to solve the aforementioned issues. Efficient inspection bit prediction models for unbiased response extraction are proposed and validated. Our experimental results of the strong UNBIAS PUF show 5.9\% intra-Fractional Hamming Distance (FHD) and 45.1\% inter-FHD on 7 Field Programmable Gate Array (FPGA) boards without applying any physical layout constraints or additional XOR gates. The UNBIAS PUF is also scalable because no characterization cost is required for each challenge to compensate the implementation bias. The averaged intra-FHD measured at worst temperature and voltage variation conditions is 12\%, which is still below the margin of practical Error Correction Code (ECC) with error reduction techniques for PUFs.



\end{abstract}

\section{Introduction}
Hardware security has become an important aspect in modern Integrated Circuit (IC) design industry because of the global supply chain business model. Identifying and authenticating each fabricated components of a chip is a challenging task \cite{Hussain14}. A Physical Unclonable Function (PUF) has been a promising security primitive such that its behavior, or Challenge Response Pair (CRP) \cite{Roel2010}, is uniquely defined and is hard to predict or replicate. A PUF can enable low overhead hardware identification, tracing, and authentication during the global manufacturing chain. 


Silicon delay based strong PUFs have been studied intensively since its first appearance in \cite{Blaise02} because of its low implementation cost and large CRP space compared with a weak PUF \cite{Herder2014}. However, there are still design challenges that restrain a strong PUF from being put in a widespread practical use. One of the major design challenges for a silicon delay based PUF is the strict symmetric delay path layout requirement. The wire delays of the competing paths should be designed and matched carefully to avoid biased responses, otherwise low inter-chip uniqueness would make the PUF unusable \cite{Daihyun2005,Sahoo2015}. In addition to asymmetric routing, another source of the biased responses for silicon based PUF is the systematic process variation, which can also degrade the quality of a PUF, such as uniqueness or unpredictability. Finally, the metastability issue of the arbiter circuit for an Arbiter PUF can cause unstable PUF responses, making a portion of the CRP unusable due to their instabilities \cite{PotkonjakM14}.

\subsection{Asymmetric Path Delay Routing} \label{section:r_works}

For a delay based PUF, the randomness should be contributed only by the subtle variations between devices, so having biased delay differences due to asymmetric routing is detrimental to delay based PUFs, and such impact should be eliminated. However, a precise control of the routing can be a difficult and time consuming task.

An implementation of an Arbiter PUF on Field Programmable Gate Array (FPGA) is considered much more difficult than a RO PUF because the connections to the arbiter circuit must also be symmetric \cite{Prasad2016}, and performing completely symmetric routing is physically infeasible in most cases \cite{Sahoo2015}, resulting small inter-Fractional Hamming Distance (FHD) for an Arbiter PUF \cite{maiti2013}. One of the most common solutions to the asymmetric routing is to use hard-macros in FPGA designs \cite{Maiti2009,Chongyan2015}, but it is not effective with Arbiter PUF, and some less commonly-used features of the FPGA would be required \cite{Morozov10}. Other approaches try to extract randomness by XORing the outputs of multiple Arbiter PUFs at the cost of large hardware overhead and less stability \cite{Machida2015}. In \cite{Kodýtek15}, the authors proposed using `middle' bits instead of the Most Significant Bit (MSB) as the RO PUF response measurement. The measurement can effectively eliminate the biased responses, but an efficient way of predicting the inspection bit is not described, and the presented RO PUF is not a strong PUF. A RTL-based PUF bit generation unit was proposed in \cite{Anderson2010}, but to the best of our knowledge, a strong PUF that can be implemented efficiently without any layout constraints has not yet been proposed. 


\subsection{Systematic Process Variation}
The existence of systematic process variation can degrade the quality of silicon based PUFs because the local randomness should be the only desired entropy source of the delay based PUF \cite{Chi-En13}. The effect of systematic process variation is similar to having biased wire delay between two delay paths, which can also damage the uniqueness of the PUF. Another possible vulnerability caused by systematic variation is the induced process side channel attack as described in \cite{Wang16}. Due to intra-wafer systematic variation \cite{Lerong11}, PUFs fabricated at the same region on different wafers can have similar systematic behavior, which can be exploited as a process side channel attack. 

To account for systematic variations, a compensation technique is proposed in \cite{Maiti2009}, which requires careful design decisions to compare RO pairs that are physically placed close to each other. In \cite{Chi-En13}, the systematic variation is modeled and subtracted from the PUF response to distill true randomness with the cost of model calculation. Similarly, in \cite{Feiten15}, the averaged RO frequency is subtracted from the original frequency, where the multiple measurements of each RO can lead to large latency overhead. In \cite{Zhang15}, a method is proposed to extract local random process variation from total variation, however, a second order difference calculation is needed, and hard-macro technique must be applied to construct symmetric delay paths.

\subsection{Metastability of the Arbiter Circuit}
The idea of an Arbiter PUF is to introduce a race condition on two paths and an arbiter circuit is used to decide which one of the two paths reached first. The arbiter circuit is usually a D flip-flop or a SR latch. If two signals arrive at an arbiter within a short time, the arbiter circuit may enter a metastable state due to setup time violation \cite{Portmann95}. Once the arbiter circuit is in metastable state, the response becomes unstable. To eliminate the inconsistency caused by metastability of the arbiter circuit, existing approaches use the majority vote or to choose the paths that have a delay difference larger than the metastable window $\delta$ at the cost of CRP characterization and discarding the unstable CRPs \cite{PotkonjakM14}.

\subsection{Our Contributions}
In this paper, we propose the physical implementation bias agnostic (UNBIAS) PUF that is immune to physical implementation bias. The contributions of this paper include:
\begin{itemize}
\item We proposed the first strong UNBIAS PUF that can be implemented purely by RTL without imposing any physical routing constraints.
\item Efficient inspection bit selection strategy based on intra-/inter-FHD prediction models are proposed and verified on the strong UNBIAS PUF.

\end{itemize}


\section{Proposed Strong UNBIAS PUF} \label{sec:bias_puf}
\begin{figure}[htb]
\centering
\includegraphics[width=3.3in]{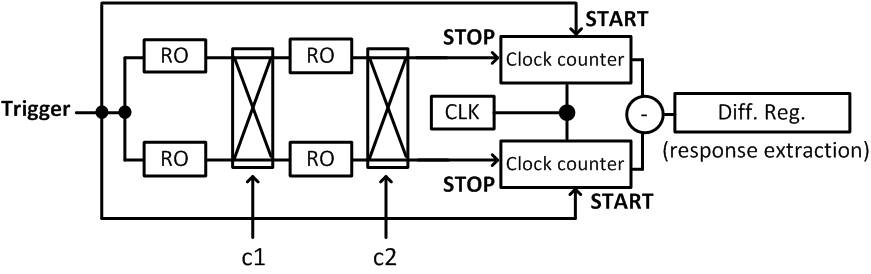}
\caption{The proposed strong UNBIAS PUF. The Clock counter starts counting clock cycles of the system clock (CLK) when START arrives and stops when STOP arrives. The difference of two Clock counters are stored in the difference register for further response extraction.}
\label{figure:biaspuf}
\end{figure}

The proposed strong UNBIAS PUF compares two delay paths to generate PUF responses. Similar to Arbiter PUF, each bit of the challenge of the UNBIAS PUF specifies the path configuration of the delay path. As shown in Figure \ref{figure:biaspuf}, the challenge c1 and c2 specify the path configurations, and an one-bit response is extracted from the difference register, which can be of several bits long. Once a challenge is given, a signal is applied at Trigger. Each of the Clock counter begins to count the number of clock cycles of the system clock (CLK) whenever the the signal from Trigger is propagated to the START input of the counter, and stops counting whenever the the signal from Trigger is propagated to the STOP input of the counter. For each challenge, the difference value of the two Clock counters is stored in the difference register for further response extraction, which is described in details in Section \ref{sec:measurement}.

\begin{figure}[htb]
\centering
\includegraphics[width=3in]{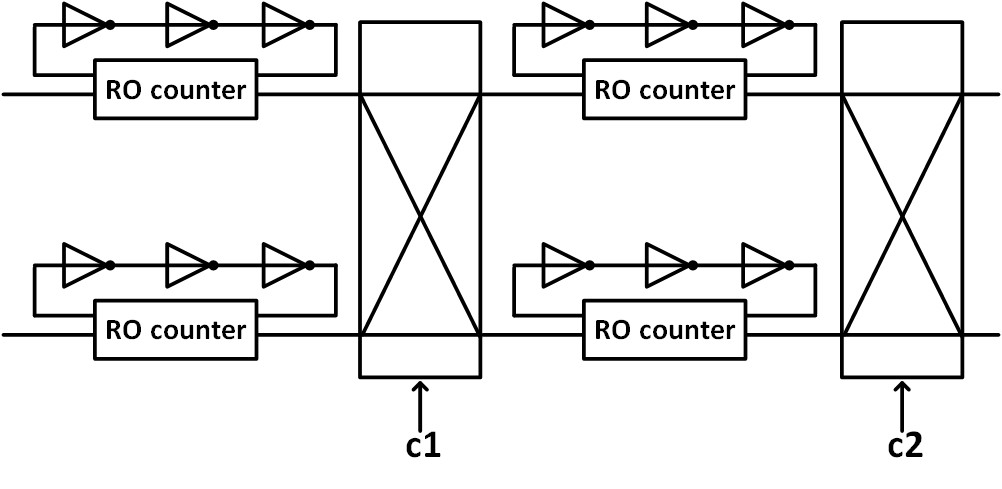}
\caption{ROs are inserted between path configurations to increase the path delay for better stability. The signal from previous path configuration is propagated only when the count of the RO counter reaches a certain threshold.}
\label{figure:ro_delay}
\end{figure}

The purpose of the ROs inserted between path configurations is to increase the path delay so that it will take multiple clock cycles for the signal to propagate to stop the clock counter. As shown in Figure \ref{figure:ro_delay}, each RO is associated with a RO counter that counts the number of oscillations of the RO. The RO counter starts counting when the signal from its previous path configuration is arrived, and propagates the signal to the next path configuration only when the count reaches a certain threshold. All the ROs are composed of same number of inverters and neither configurations nor any layout constraints are needed.

Unlike the conventional Arbiter PUF, the strong UNBIAS PUF has no metastability issues caused by a D flip-flop or a latch. The delay difference of the two paths is transformed into counter values of the system clock. By judiciously extracting the response from the difference register, the physical implementation bias can be effectively mitigated, therefore the UNBIAS PUF can be implemented purely by RTL without any routing or layout constraints. Details of the response extraction are described in Section \ref{sec:measurement}.

\section{Bias-Immune Response Extraction} \label{sec:measurement}


\subsection{Inspection Bit on Unbiased/Biased Paths} \label{sec:ins}
\begin{figure}[htb]
\centering
\includegraphics[width=2.8in]{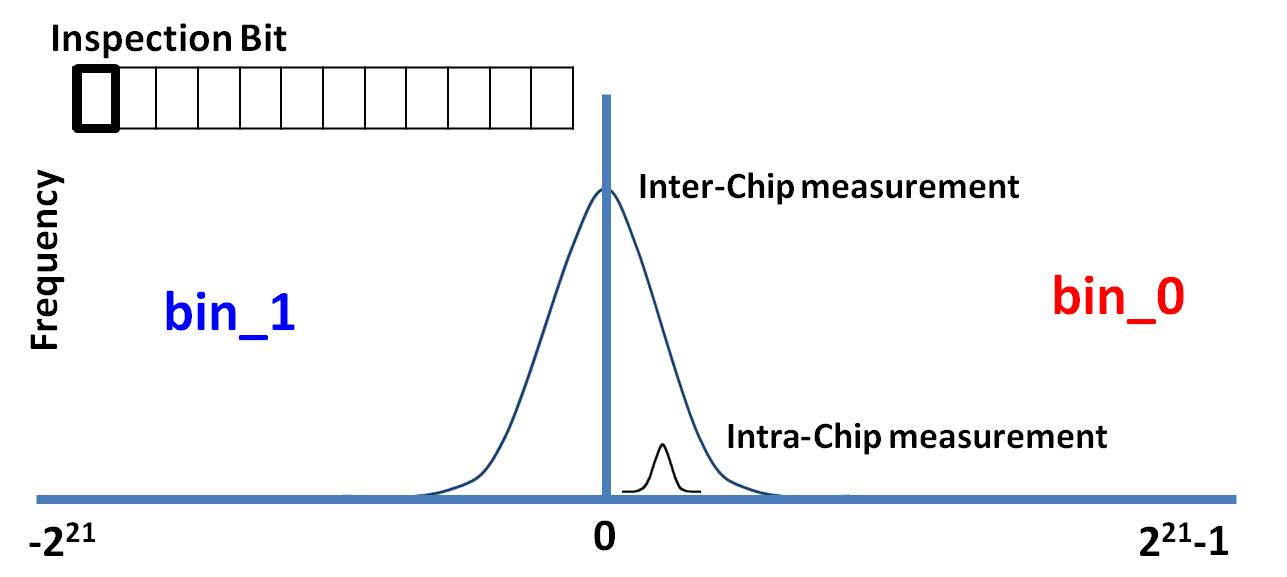}
\caption{For a symmetrically routed PUF, the inter-FHD would be close to 50\%. The intra-FHD may not be zero due to measurement noise.}
\label{figure:ud}
\end{figure}

In this section we describe how different selections of the inspection bit can change the intra- and inter-FHD. Figure \ref{figure:ud} shows an example of a distribution of values from difference registers of symmetrically routed UNBIAS PUFs. The length of the difference register is 22-bit, so the range of the register value is between $-2^{21}$ and $2^{21}-1$ as represented in 2's-complement. The large inter-chip measurement curve gives the distribution of the values across all PUFs. Since the PUF is unbiased, roughly half of the difference values would be greater than zero due to random local process variation, therefore the inter-FHD of the UNBIAS PUFs would be close to 50\%. In this case, the inspection bit is simply the MSB, which divides the range of 22-bit difference value into two groups $bin\_1$ and $bin\_0$. All measurements fall into $bin\_1$ on the left output a 1; others output a 0. The small intra-chip measurement curve gives the distribution of multiple measurements of the PUF on a same chip. Due to noise, the difference values could be different, so the intra-FHD of the difference register may not be a perfect 0\%.

\begin{figure}[htb]
\centering
\includegraphics[width=2.9in]{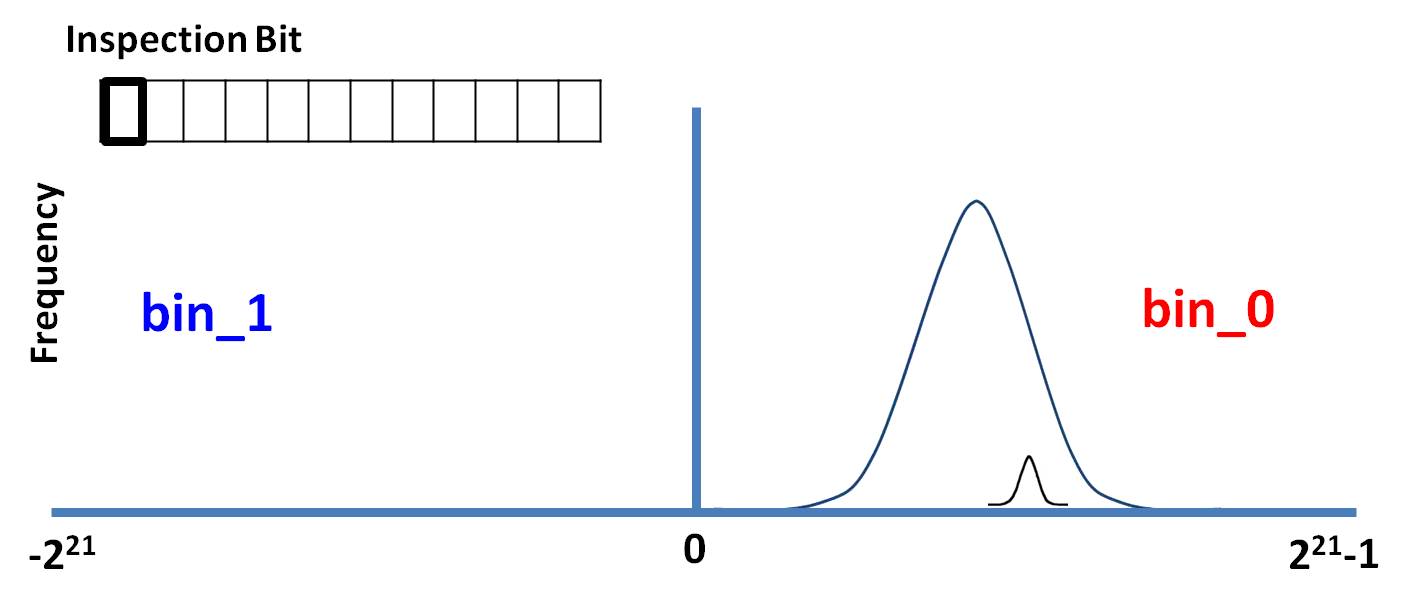}
\caption{For a biased PUF, most of the difference values across all chips could be greater than zero, causing a low inter-FHD if the MSB is the inspection bit.}
\label{figure:bd}
\end{figure}

Even though symmetric UNBIAS PUF layout is much preferred, it is difficult and takes much effort and overhead to achieve such requirement as described in Section \ref{section:r_works}. In practice, if no layout constraints are imposed, the measurement distribution of the difference register can be as shown in Figure \ref{figure:bd}, where most of the difference values across chips are greater than zero. In this case, using the MSB as the inspection bit would cause low inter-FHD of the PUFs because most MSBs are 0's.

\begin{figure}[htb]
\centering
\includegraphics[width=2.9in]{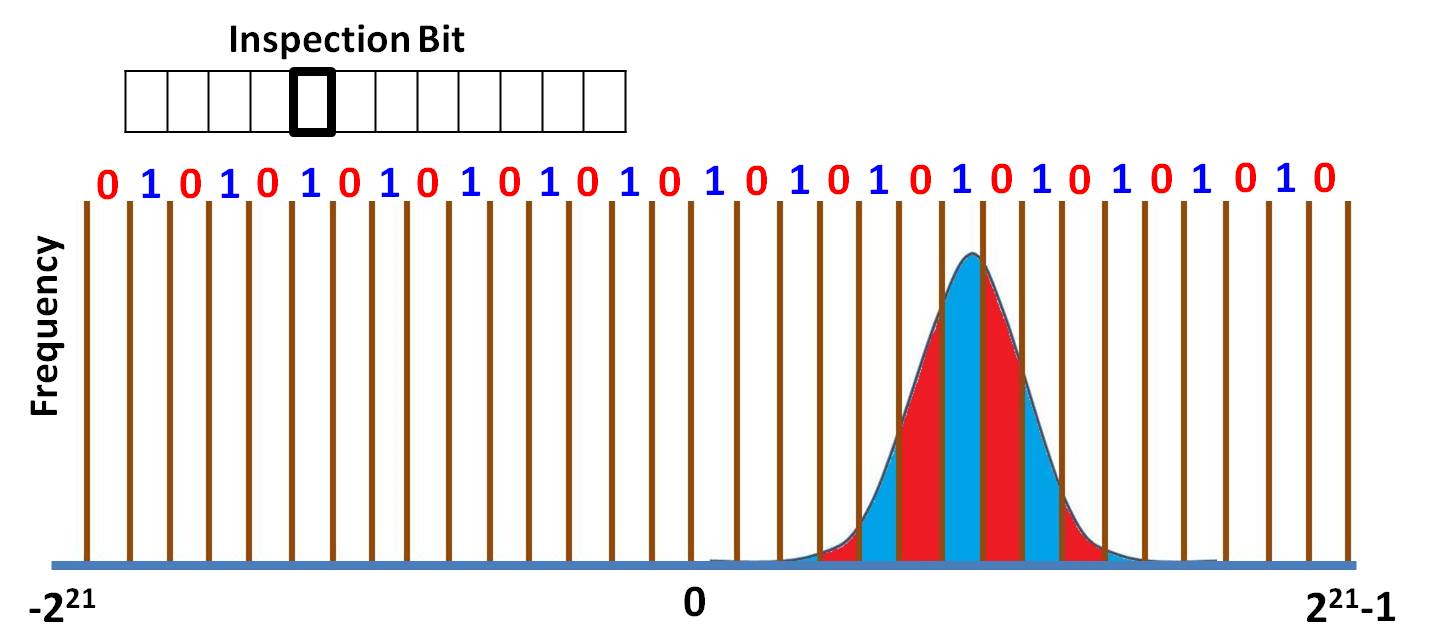}
\caption{For an asymmetrically routed PUF with proper inspection bit, roughly half of the difference values across all chips would fall in $bin\_1$, therefore the inter-FHD would be close to 50\%.}
\label{figure:bd_bin}
\end{figure}

For the same biased distribution shown in Figure \ref{figure:bd}, if the $i^{th}$ bit is used as the inspection bit of the difference register as Figure \ref{figure:bd_bin} shows, the range of the 22-bit difference value is divided into multiple bins with width $2^{i}$, where the output of the measurement is decided by the bin in which it resides. Note that in this case the response is not an indicator of which delay is longer in the comparison. The smaller the width of the bin is, the closer the inter-FHD is to 50\% because roughly half of the outputs would reside in $bin\_1$ even with biased delay. On the other hand, the width of the bin should be large enough so that multiple measurements of a same PUF should always fall into the same bin. In other words, the width of the bin should be larger than the variation of the intra-chip measurement distribution. Therefore, the choice of inspection bit is a tradeoff between inter-FHD and intra-FHD for a PUF with asymmetric routing.

\section{Inspection Bit Identification} \label{sec:models}

\subsection{Intra-FHD Prediction Model}  \label{sec:up}
The intra-FHD depends on the width of the bins $w=2^{i}$ when the inspection bit is $bit_{i}$. A straightforward way to determine the associated intra-FHD for each inspection location is to gather multiple measurements of the same challenge on a same PUF, and simply calculate the intra-FHD for each $bit_{i}$. A more efficient approach is to predict the intra-FHD without calculating it for each $bit_{i}$. 

To predict intra-FHD$_{k}$ of a challenge $C_{k}$ of an inspection bit, we first obtain $t$ measured difference registers of the challenge $C_{k}$ of a same PUF. Since the bin width and the range of the difference register is known, the $t$ difference values can be divided into two groups (responses) according to the bins they reside in. Let the number of difference values fall in $bin\_1$ be $n_{one}$, and number of difference values fall in $bin\_0$ be $n_{zero}$. $n_{one}$ and $n_{zero}$ represent the number of responses of the challenge $C_{k}$ to be one and zero during the $t$ measurements, respectively. Since the intra-FHD is essentially calculated from the response difference between any two measurements, the predicted intra-FHD$_{k}$ is calculated as:


\begin{equation} \label{eq:ratio}
\begin{aligned}
\textit{intra-FHD}_{k} &= \frac{n_{one}\times n_{zero}}{\binom{t}{2}} \times 100\%,
\end{aligned}
\end{equation}

where the final predicted intra-FHD would be the averaged intra-FHD$_{k}$ of all challenges.

\begin{figure}[htb]
\centering
\includegraphics[width=2.7in]{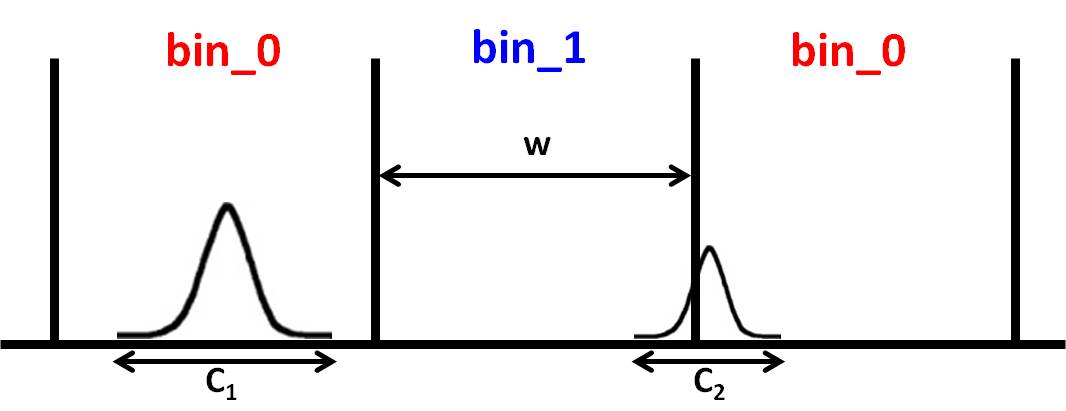}
\caption{Magnified view of Figure \ref{figure:bd_bin} with three bins. $w$ is the bin width and the measurement ranges for challenges $C_{1}$ and $C_{2}$ are specified. The expected intra-FHD$_{1}$ is 0\% and the expected intra-FHD$_{2}$ depends on the portion of measured values that fall in $bin\_1$.}
\label{figure:vw}
\end{figure}

As shown in Figure \ref{figure:vw}, the expected intra-FHD$_{1}$ is 0\% because all measurements fall in the same bin and $n_{one} \times n_{zero} =0$. The expected intra-FHD$_{2}$ depends on the portion of measured values that fall in $bin\_1$. With larger bin width $w$, it is more likely that all responses would fall into the same bin


\subsection{Inter-FHD Lower Bound Prediction Model}
The inter-FHD depends on the bin width $w$ with a given inspection bit $bit_{i}$. 
Assume the distribution of inter-chip difference value is a Normal distribution $N\sim(\mu,\sigma^2)$. Define $\epsilon$ to be the distance between the mean $\mu$ and the closest bin boundary on the left as Figure \ref{figure:iwc} shows. We first prove that the worst-case inter-FHD happens when $\epsilon=0.5w$, followed by the prediction model of the inter-FHD for the worst-case scenario. 

\subsubsection{Worst-Case Inter-FHD Identification}
Given a fixed $w$, define $A_{1}(\epsilon)$ and $A_{0}(\epsilon)$ to be the total underlying area in $bin\_1$ and $bin\_0$ as functions of $\epsilon$, respectively. For any Normal distribution, $A_{1}(\epsilon)$ and $A_{0}(\epsilon)$ are calculated as:

\vspace{-5mm}
\begin{flalign}
A_{1}(\epsilon) &= \sum_{n=-\infty}^{\infty} F(-\epsilon+2nw+w)-F(-\epsilon+2nw) \label{eq:A1} \\ 
A_{0}(\epsilon) &= 1 - A_{1}(\epsilon,w) \label{eq:A0}
\end{flalign}
where $F(\cdot)$ is the Cumulative Distribution Function (CDF) of the Normal distribution, and $n$ is the index for bin area summation. 

The ratio $Ratio(\epsilon)$ is defined as:
\begin{equation} \label{eq:ratio}
\begin{aligned}
Ratio(\epsilon) &= \frac{A_{1}(\epsilon)}{A_{0}(\epsilon)},                      \quad 0<\epsilon<w\\
\end{aligned}
\end{equation}
where the range of $\epsilon$ is from 0 to $w$ because of its periodic structure.

\begin{figure}[htb]
\centering
\includegraphics[width=2.2in]{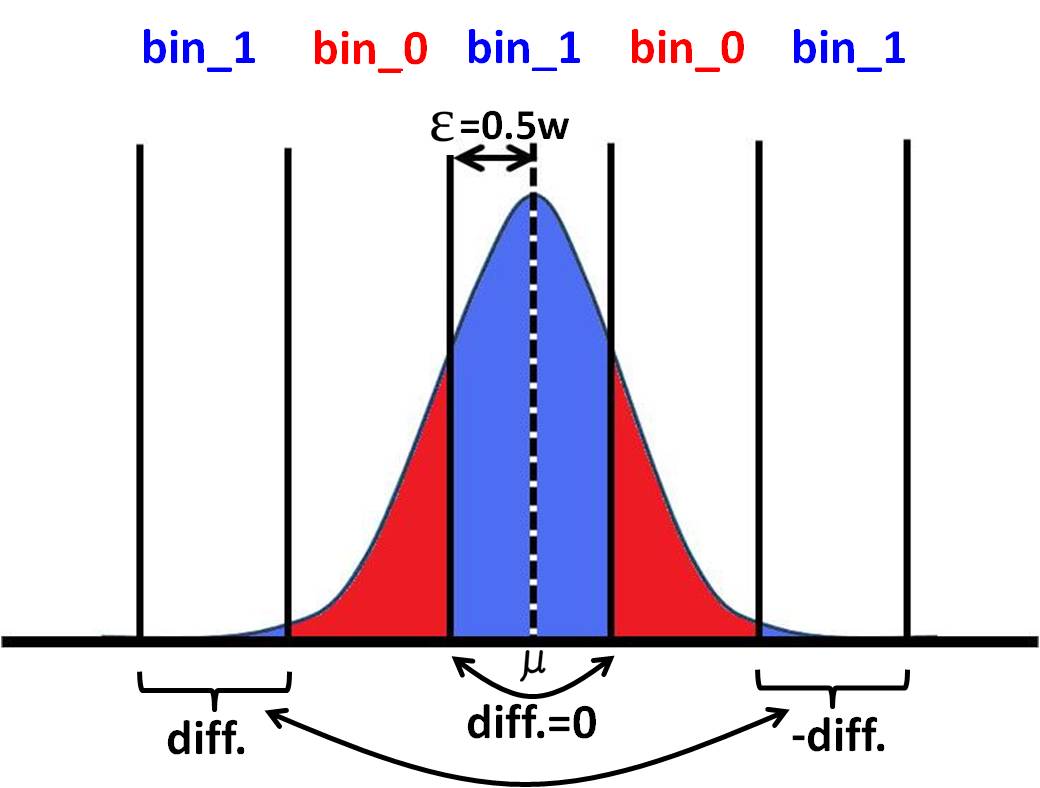}
\caption{Worst Inter-FHD happens when the mean is at the middle of a bin.}
\label{figure:iwc}
\end{figure}

The closer the $Ratio(\epsilon)$ is to one, the closer the inter-FHD would be to 50\% because the two areas are closer to each other. We want to show that the largest (most unbalanced) ratio happens at $\epsilon=0.5w$ as Figure \ref{figure:iwc} shows. 

To find the extreme value of $Ratio(\epsilon)$ given a fixed $w$, we take derivative with respective to $\epsilon$ of Equation \ref{eq:ratio} and replace $A_{0}(\epsilon)$ by $1-A_{1}(\epsilon)$ from Equation \ref{eq:A0}: 
\begin{equation} \label{eq:ratio_de}
\begin{aligned}
\frac{d}{d\epsilon}Ratio(\epsilon) &= \frac{A'_{1}(\epsilon)}{(1-A_{1}(\epsilon))^{2}}
\end{aligned}
\end{equation}

From Equation \ref{eq:ratio_de} we see that to find the extreme value of $Ratio(\epsilon)$, it is equivalent to find the solution of $A'_{1}(\epsilon)$, which is given below:
\begin{equation} \label{eq:A1_de}
\begin{aligned}
\frac{d}{d\epsilon}A_{1}(\epsilon) &= \sum_{n=-\infty}^{\infty} f(-\epsilon+2nw+w)-f(-\epsilon+2nw) \\
\end{aligned}
\end{equation}
where $f(\cdot)$ is the Probability Density Function (PDF) of the Normal distribution. Equation \ref{eq:A1_de} shows that $A'_{1}(\epsilon)$ is the summation of differences between two PDF terms where one is a shifted version by $w$ of another. Therefore, applying $\epsilon=0.5w$ to Equation \ref{eq:A1_de}, we get a zero. Figure \ref{figure:iwc} shows that when $\epsilon=0.5w$, each difference term in Equation \ref{eq:A1_de} has its counter part at the mirrored location to the center, so that the summation becomes zero.

To conclude our derivation, given a $w$ of an inspection bit, the extreme value of $Ratio(\epsilon)$ happens when $\epsilon=0.5w$, and the inter-chip stander deviation $\sigma$ is needed for the $Ratio(\epsilon)$ calculation.

\subsubsection{Inter-FHD Lower Bound Prediction}
To predict inter-FHD, we calculate the probability of which any pair of chips produce different responses. The inter-FHD prediction given the width $w$ of the inspection bit is:
\begin{equation} \label{eq:iner}
\begin{aligned}
\textit{inter-FHD}	&=\frac{2Ratio(\epsilon)}{(1+Ratio(\epsilon))^{2}}  \\
\end{aligned}
\end{equation}

With $Ratio(\epsilon)=1$, the two areas are the same, resulting a predicted 50\% inter-FHD. Given a selected $bit\_i$, plugging in $\epsilon=0.5w$ to Equation \ref{eq:iner} would give the predicted inter-FHD lower bound.

Please note that to predict inter-FHD, the inter-chip standard deviation $\sigma$ is needed because the calculation involves the CDF. However, the mean $\mu$ does not affect the prediction because the extreme value is obtained by finding the worst-case $\epsilon$. Also, since changing the inspection bit results at least a 2x change of $w$, the inter-chip $\sigma$ does not have to be calculated with high accuracy. It can be obtained by pre-layout simulation or measuring a small number of chips.

\subsection{Inspection Bit Selection}
Given the Error Correction Code (ECC) specification corresponding to the PUF design, the intra-FHD threshold can be defined. 
From the intra-FHD prediction model, choose a set of candidate bits that would satisfy the intra-FHD threshold requirement. From the candidate bits, a best inspection bit can be determined by applying the inter-FHD prediction model given the standard deviation $\sigma$ of the inter-chip delay distribution.

Please note that only one chip is needed for the inspection bit selection since the measurement noise is similar for all chips from our experiment and the $\sigma$ is obtained from pre-layout simulation. The location of the final inspection bit, which is a public information, is passed to all PUFs for the secret response generation.

\section{Experimental Results} \label{sec:exp}

\subsection{Strong UNBIAS PUF Implementation}
The strong UNBIAS PUF structure is implemented on 7 Altera DE2-115 FPGA boards. In our implementation, no physical constraints, additional XORs, tunable delay units, or any systematic variation compensation techniques are used. The design is purely a RTL design.

The ROs inserted between path configurations are composed of 19 inverters, and the signal will be propagated to the next path configuration when the RO counter associated to the RO reaches a count of 50 thousand. The UNBIAS PUF has 10 path configurations, therefore the length of the challenge is 10-bit long. The length of the difference register is 19-bit, and the length of the final response for each challenge is one bit. For our experiment, 120 challenges are applied, and 120 bits of responses are obtained for each PUF within a second. Please note that the RO structure and the count of the RO counter are selected given the 50 MHz system clock of the FPGA. The results are similar as long as no overflow occurs at the 19-bit difference register.


\subsection{Prediction Model Validation}
The inter-FHD is obtained from 7 FPGAs, and the intra-FHD is calculated by measuring each PUF 10 times. To show inter-chip variation and measurement noise of our experimental setup, we measure the frequency of a single RO across the chips 10 times, and the inter-chip variation is 6.1\% with 0.2\% measurement noise.

\begin{figure}[h]
\centering
\includegraphics[width=2.7in]{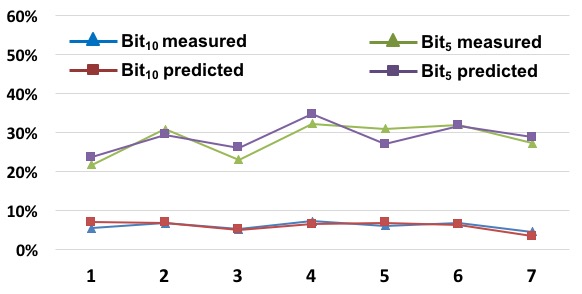}
\caption{Strong UNBIAS PUF intra-FHD predictions of $bit_{5}$ and  $bit_{10}$ of 7 FPGAs. $bit_{5}$ has much larger intra-FHD because its bin width is smaller.}
\label{figure:intra_predict}
\end{figure}

To validate the intra-FHD prediction model, we follow the procedure described in Section \ref{sec:up} with $t=10$ measurements. Figure \ref{figure:intra_predict} shows the results of the intra-FHD prediction of  $bit_{5}$ and  $bit_{10}$. The intra-FHD of $bit_{5}$ is much higher than  $bit_{10}$ because its bin width is much smaller. 

To validate the inter-FHD prediction model, for each challenge, we obtain an inter-chip standard deviation $\sigma$ from 7 FPGAs, and the final $\sigma$ used in the prediction model is the median of the $\sigma$ from 120 challenges, which gives $\sigma=521$. The results shown in Figure \ref{figure:inter_predict} indicate that the inter-FHD lower bound prediction is well matched with the measured data. 
To demonstrate that the inter-FHD prediction model does not require an accurate inter-chip $\sigma$ estimation, Figure \ref{figure:inter_predict} also shows the prediction range with $\sigma \pm 15\%$ variation. We can see that the differences of the predictions are limited, which indicates that the $\sigma$ can either be obtained from pre-layout simulation or measurements of a small number of chips. The prediction gap is relatively large when $w$ is much larger than $\sigma$. However, as $w$ becomes comparable to $\sigma$, where potential inspection bits begin to occur, the prediction curve rises up quickly and matches the measured data well. Figure \ref{figure:inter_predict} also shows that $bit_{10}$ should be a proper inspection bit because the intra-FHD is low and the inter-FHD is close to 45\%

\subsection{Uniqueness and Reliability Evaluation} \label{sec:measure}

\begin{figure}[htb]
\centering
\includegraphics[width=2.9in]{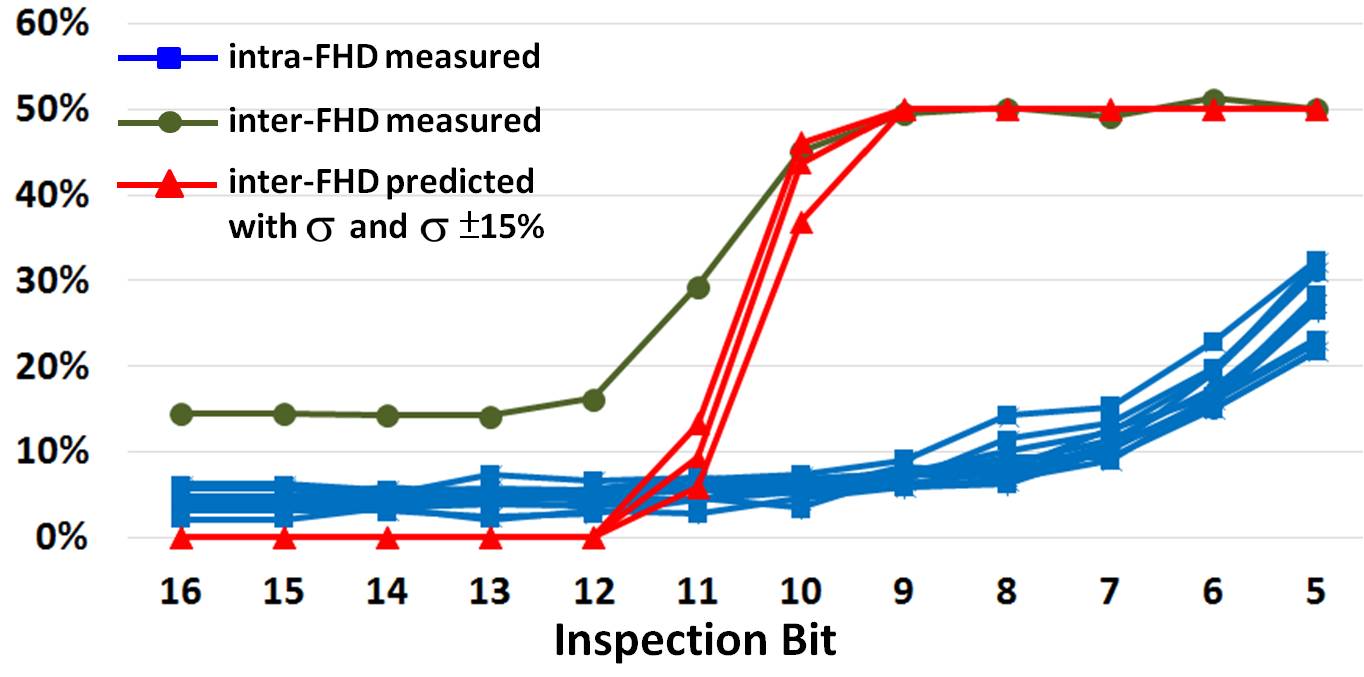}
\caption{Inter-, intra-FHD, and inter-FHD prediction using $\sigma=521$ with different inspection bit selections of the strong UNBIAS PUF.}
\label{figure:inter_predict}
\end{figure}

The results of inter-FHD and intra-FHD with different inspection bit selections are shown in Figure \ref{figure:inter_predict}. As we can see from the figure, using bits closer to the MSB gives low intra-FHD but also low inter-FHD. This verifies the fact that the delay paths are biased if no physical implementation constraints are imposed. On the other hand, using bits closer to the LSB gives 50\% on both intra-FHD and inter-FHD because of the measurement noise. As predicted, the best inspection location appears at $bit_{10}$ with 45.1\% inter-FHD and 5.9\% intra-FHD. The results also indicate that the systematic variation is mitigated because no constraints are imposed at all.

%

Table \ref{table:compare} shows comparison results with previous work. With conventional Arbiter PUF (APUF) shown in the second column, the results from \cite{maiti2013} show that the circuit is essentially a constant number generator with very little inter-FHD. The third column shows the 3-1 double Arbiter PUF with XORs \cite{Machida2015}, where symmetric layout is still required, and the hardware overhead is 2X or 3X from the duplicated circuits depending on the uniqueness requirement of the application. The inter-FHD is close to 50\% but the intra-FHD is high due to the XORs. The fourth column shows the results from Path Delay Line (PDL) PUF \cite{Sahoo2015}. Symmetric PDL and delay characterization for each CRP are required, which can cause scalability issues. Also the ability of eliminating biased responses is limited because it depends on the number of tuning stages inserted. The last column shows the proposed strong UNBIAS PUF. Its behavior is unique and stable, and most importantly no symmetric layout at all. 


\begin{table}
\centering
\caption{Comparison between previous Arbiter PUFs and strong UNBIAS PUF}
\begin{tabular}{|c|c|c|c|c|}
\hline
 & \makecell{APUF\\ \cite{maiti2013}} &  \makecell{XOR\\ \cite{Machida2015}} 	&  \makecell{PDL\\ \cite{Sahoo2015}}	& \makecell{UNBIAS\\PUF}	\\
\hline
inter-FHD			& 7.2\% & 50.6\% & 45.25\%		& 45.1\% \\
\hline
intra-FHD			& 0.24\% & 11.8\% & 4.1\%		& 5.9\% \\
\hline
Symm. Layout			& No & Yes & Yes		& No \\
\hline
Characterization			& No & No & Yes		& No \\
\hline
\end{tabular}
\label{table:compare}
\end{table}

\subsection{Temperature and Voltage Variations}
For temperature and voltage variations, the reference responses are measured at 20$^{\circ}$C with standard voltage 12V. The reference responses are then compared with responses measured at 20$^{\circ}$C and 75$^{\circ}$C with 10\% voltage variation. The results indicate the reliability of the PUF when it is enrolled at normal condition but verified at a high temperature environment with unstable voltage source.

\begin{figure}[h]
\centering
\includegraphics[width=2.4in]{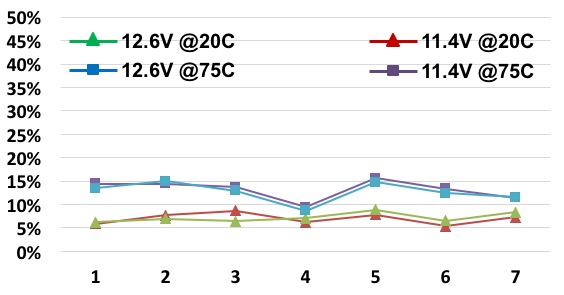}
\caption{Strong UNBIAS PUF intra-FHD under tempreature and voltage variations.}
\label{figure:variations}
\end{figure}

Figure \ref{figure:variations} shows the intra-FHD using $bit_{10}$ as the inspection bit. All intra-FHD at 20$^{\circ}$C with 10\% voltage variation is below 8\%, and all intra-FHD at 75$^{\circ}$C with 10\% voltage variation is below 14\%, which is still within conventional ECC margin with error reduction techniques for PUFs \cite{Guajardo07,Yu2010}. Compared with RO PUF presented in \cite{Kodýtek15}, one possible explanation of smaller intra-FHD for our strong UNBIAS PUF is that with multiple RO delay units, the overall delay variation is canceled out, where for the RO PUF, the variation of each RO is directly compared.

\vspace{-5mm}

\section{Conclusions} \label{sec:con}
We proposed the first strong UNBIAS PUF that can be implemented purely by RTL without complex post-layout analysis or hand-crafted physical design effort. The proposed measurement can effectively mitigate the impact of biased delay paths and metastability issues to extract local device randomness. The inspection bit can be determined efficiently from the intra-FHD and inter-FHD prediction models. 

The strong UNBIAS PUF is implemented on 7 FPGAs without imposing any physical layout constraints. Experimental results show that the intra-FHD of the strong UNBIAS PUF is 5.9\% and the inter-FHD is 45.1\%, and the prediction models are closely fitted to the measured data. The averaged intra-FHD of the strong UNBIAS PUF at worst temperature and voltage variations is about 12\%, which is still within the margin of conventional ECC techniques. The fact that the proposed scheme is immune to physical implementation bias would allow the strong UNBIAS PUF to be designed and integrated with minimum effort in a high-level description of the design, such as during RTL design.

\footnotesize{
\bibliographystyle{unsrt}
\bibliography{PUF} 

\begin{thebibliography}{10}

\bibitem{Hussain14}
S.~U. Hussain, S.~Yellapantula, M.~Majzoobi, and F.~Koushanfar.
\newblock {BIST-PUF: Online, Hardware-based Evaluation of Physically Unclonable
  Circuit Identifiers}.
\newblock In {\em Proc. ICCAD}, 2014.

\bibitem{Roel2010}
Roel Maes and Ingrid Verbauwhede.
\newblock {Physically Unclonable Functions: A Study on the State of the Art and
  Future Research Directions}.
\newblock In {\em {Towards Hardware-Intrinsic Security}}, pages 3--37.
  {Springer Berlin Heidelberg}, 2010.

\bibitem{Blaise02}
Blaise Gassend et~al.
\newblock {Silicon Physical Random Functions}.
\newblock In {\em Proc. CCSC}, 2002.

\bibitem{Herder2014}
C.~Herder et~al.
\newblock {Physical Unclonable Functions and Applications: A Tutorial}.
\newblock {\em Proc. of the IEEE}, pages 1126--1141, Aug 2014.

\bibitem{Daihyun2005}
Daihyun Lim et~al.
\newblock {Extracting secret keys from integrated circuits}.
\newblock {\em IEEE Transactions on VLSI Systems}, 2005.

\bibitem{Sahoo2015}
D.~P. Sahoo, R.~S. Chakraborty, and D.~Mukhopadhyay.
\newblock {Towards Ideal Arbiter PUF Design on Xilinx FPGA: A Practitioner's
  Perspective}.
\newblock In {\em Euromicro Conference on DSD}, 2015.

\bibitem{PotkonjakM14}
Teng Xu and M.~Potkonjak.
\newblock {Robust and flexible FPGA-based digital PUF}.
\newblock In {\em International Conference on FPL}, 2014.

\bibitem{Prasad2016}
Durga~Prasad Sahoo et~al.
\newblock {Architectural Bias: a Novel Statistical Metric to Evaluate Arbiter
  PUF Variants}.
\newblock Cryptology ePrint Archive, Report 2016/057, 2016.

\bibitem{maiti2013}
V.~Gunreddy A.~Maiti and P.~Schaumont.
\newblock {A Systematic Method to Evaluate and Compare the Performance of
  Physical Unclonable Functions}.
\newblock In {\em Embedded Systems Design with FPGAs}, 2013.

\bibitem{Maiti2009}
A.~Maiti and P.~Schaumont.
\newblock {Improving the Quality of a Physical Unclonable Function using
  Configurable Ring Oscillators}.
\newblock In {\em International Conference on FPL}, 2009.

\bibitem{Chongyan2015}
Chongyan Gu and M.~O'Neill.
\newblock {Ultra-compact and Robust FPGA-based PUF Identification Generator}.
\newblock In {\em IEEE ISCAS}, 2015.

\bibitem{Morozov10}
S.~Morozov et~al.
\newblock {An analysis of delay based PUF implementations on FPGA}.
\newblock In {\em {Reconfigurable Computing: Architectures, Tools and
  Applications}}. {Springer Berlin Heidelberg}, 2010.

\bibitem{Machida2015}
Mitsugu~Iwamoto Takanori~Machida, Dai~Yamamoto and Kazuo Sakiyama.
\newblock {A New Arbiter PUF for Enhancing Unpredictability on FPGA}.
\newblock In {\em The Scientific World Journal}, 2015.

\bibitem{Kodýtek15}
F.~Kodýtek and R.~Lórencz.
\newblock {A Design of Ring Oscillator Based PUF on FPGA}.
\newblock In {\em IEEE International Symposium on DDECS}, 2015.

\bibitem{Anderson2010}
J.~H. Anderson.
\newblock {A PUF design for secure FPGA-based embedded systems}.
\newblock In {\em Proc. ASP-DAC}, 2010.

\bibitem{Chi-En13}
Chi-En Yin and Gang Qu.
\newblock {Improving PUF Security With Regression-based Distiller}.
\newblock In {\em Proc. DAC}, 2013.

\bibitem{Wang16}
Wei-Che Wang, Yair Yona, Suhas Diggavi, and Puneet Gupta.
\newblock {LEDPUF: Stability-Guaranteed Physical Unclonable. Functions through
  Locally Enhanced Defectivity}.
\newblock In {\em IEEE International Symposium on HOST}, 2013.

\bibitem{Lerong11}
Lerong Cheng et~al.
\newblock {Physically Justifiable Die-Level Modeling of Spatial Variation in
  View of Systematic Across Wafer Variability}.
\newblock {\em IEEE TCAD}, 2011.

\bibitem{Feiten15}
Linus Feiten et~al.
\newblock {Improving RO-PUF Quality on FPGAs by Incorporating Design-Dependent
  Frequency Biases}.
\newblock {\em IEEE ETS}, 2015.

\bibitem{Zhang15}
Qinglong Zhang et~al.
\newblock {FROPUF: How to Extract More Entropy from Two Ring Oscillators in
  FPGA-Based PUFs}.
\newblock {\em {IACR}}, 2015.

\bibitem{Portmann95}
L.~Portmann and Teresa H.-Y. Meng.
\newblock {Metastability in CMOS library elements in reduced supply and
  technology scaled applications}.
\newblock {\em IEEE JSSC}, 1995.

\bibitem{Guajardo07}
J.~Guajardo, G.-J.~Schrijen S.~S.~Kumar, and P.~Tuyls.
\newblock {FPGA Intrinsic PUFs and Their Use for IP Protections}.
\newblock In {\em CHES}, Sep 2007.

\bibitem{Yu2010}
M.-D.~M. Yu and S.~Devadas.
\newblock {Secure and Robust Error Correction for Physical Unclonable
  Functions}.
\newblock In {\em IEEE Des. Test}, 2010.

\end{thebibliography}
}


\end{document}